\documentclass[aps,prl,showpacs,twocolumn]{revtex4}
\usepackage{amsfonts}
\usepackage{mathrsfs}
\begin{document}
\title{Reply to the Comment of X. Ji on ``Spin and orbital angular momentum
in gauge theories'' [PRL 100:232002 (2008)]}
\author{Xiang-Song Chen,$^{1,2}$ Xiao-Fu L\"{u},$^1$ Wei-Min Sun,$^2$
Fan Wang,$^2$ and T. Goldman$^3$}
\affiliation{$^1$Department of
Physics, Sichuan University, Chengdu 610064, China\\
$^2$Department of Physics, Nanjing University, CPNPC, Nanjing
210093, China\\
$^3$Theoretical Division, Los Alamos National Laboratory, Los
Alamos, NM 87545, USA}
\date{\today}

\begin{abstract}
We reply to the Comment of X. Ji \cite{Ji08} on our paper
\cite{Chen08}, concerning angular momentum algebra, locality,
Lorentz covariance, and measurability of our gauge-invariant
description of the spin and orbital angular momentum of quarks and
gluons.
\end{abstract}
%11.15.-q Gauge field theories
%14.20.Dh Protons and neutrons
%12.38.-t Quantum chromodynamics
\maketitle

It is certainly true and common that a bare-operator algebra may
acquire scale-dependence and even anomalous terms when matrix
elements are computed, but one must start from the bare-operator
algebra when assigning appropriate physical meaning to any operator.
In fact, this is exactly what one compares to when speaking of an
anomaly. E.g., $\psi^\dagger \frac{\vec\Sigma}{2}\psi$ (but not
anything else) is identified as the quark spin operator purely
because of its bare-operator property, irregardless of the axial
anomaly. Actually, even the basic canonical quantization rule
$\{{\psi(\vec r,t),\psi^{\dag}(\vec{r'},t)}\}=\delta(\vec r-\vec
r')$ is modified by renormalization. There is no justification for
abandoning such a bare-operator algebra due to its modification in
renormalization. Moreover, these algebra are the basic elements in
quantum mechanics without renormalization, where the
gauge-invariance problem with angular momentum is already
encountered and our solution is needed.

To examine Ji's comment that our proposal ``clashes'' with locality,
one must first distinguish between locality and local expression.
Locality means vanishing of commutator at space-like intervals, 
but it by no means requires a physical expression to be local 
function of all field variables, because (in a gauge theory, 
particularly) not all field components are independent variables to 
which the quantization rules apply. It must be further clarified that 
quantum field theory can {\em accommodate} non-locality for certain 
field variables without violating causality. The point is that the
field variables are not always observables. Locality is required
only for the operators for physical observables, which must
commute at space-like intervals so that experiments at space-like
intervals produce uncorrelated results. A celebrated example is the
quantization of gauge theory in Coulomb gauge. In fact, as Strocchi
and Wightman proved in \cite{Stro74}, quantization of gauge theory
with only physical degrees of freedom necessarily involves
non-locality and complex Lorentz transformation rule.

Contrary to Ji's comment, our formalism does preserve Lorentz
covariance. To appreciate this point, recall the well-known fact
that the gauge field $A^\mu$ does {\em not}, in fact, transform as a
four-vector \cite{Bjor65,Mano87,Wein95}. Instead, it acquires an
extra pure-gauge term:
\begin{equation}
U(\Lambda) A^\mu U^{-1}(\Lambda)=\Lambda^\mu_{~\nu}A^\nu(\Lambda x)
+\partial^\mu \Omega(x,\Lambda). \label{UA}
\end{equation}
The expression for $\Omega$ can be explicitly worked out by
consistent canonical quantization and computation of the commutator
of $A^\mu$ with the generators of Lorentz transformation (a most
complete treatment is offered in \cite{Mano87}), or, in the spirit
of Weinberg \cite{Wein95}, by starting with physical photons and
constructing $A^\mu$ using the photon creation and annihilation
operators. A remarkable fact is that if one applies the apparently
non-covariant Coulomb gauge $\vec\nabla \cdot \vec A = 0$ in one
frame, then the extra pure-gauge term $\partial^\mu \Omega$ restores
$\vec\nabla '\cdot \vec A '=0$ in a transformed frame. Since the
canonical quantization rule for the physical field $\vec A_{\rm
phys}$ would be exactly the same as that for $\vec A$ in Coulomb
gauge (in which $\vec A_{\rm pure}=0$ and $\vec A_{\rm phys}=\vec
A$), in our formulation $\vec A_{\rm phys}$ transforms just as
in Eq. (\ref{UA}), and hence $\vec \nabla \cdot \vec A_{\rm phys}=0$
is preserved under Lorentz transformation. That is, the physical field
is not transformed to include non-physical components by changing
the reference frame.

Eq. (\ref{UA}) reminds us a profound fact, contrary to Ji's belief,
that it is not always possible to have all physical variables
transforming in the simple manner of Lorentz scalar, vector, etc.
Such complexity is {\em physical} and {\em intrinsic}: The
transformation rule of any operator (such as the gluon spin) is
dictated by its commutators with the generators of Lorentz
transformation, and can be explicitly worked out, though this is
certainly a non-trivial exercise in QCD.

In conclusion: the angular momentum operators which we construct are
indeed physically sound, and provide a firm and consistent basis for
further theoretical and experimental investigation of the nucleon
spin structure in terms of the four (physically intuitive)
contributions: quark/gluon spin and quark/gluon orbital angular
momentum. In the approach advocated by Ji \cite{Ji97}, however,
there is not even an identification for the gluon or photon spin,
while separate manipulation of photon spin and orbital angular
momentum is already a routine practice in modern optics
\cite{Enk07}, both theoretically and experimentally, and great
effort is being devoted to measuring the gluon spin inside the
nucleon \cite{Stra07}.

This work is supported in part by the China NSF under grants
10475057, 10875082 and 90503011, and in part by the U.S. DOE under
contract W-7405-ENG-36.

\end{document}